\documentclass[twoside]{article}

\usepackage{amsmath,amssymb}

\catcode`\@=11
\long\def\@makefntext#1{
\protect\noindent \hbox to 3.2pt {\hskip-.9pt  
$^{{\eightrm\@thefnmark}}$\hfil}#1\hfill}		

\def\thefootnote{\fnsymbol{footnote}}
\def\@makefnmark{\hbox to 0pt{$^{\@thefnmark}$\hss}}	
	
\def\ps@myheadings{\let\@mkboth\@gobbletwo
\def\@oddhead{\hbox{}
\rightmark\hfil\eightrm\thepage}   
\def\@oddfoot{}\def\@evenhead{\eightrm\thepage\hfil
\leftmark\hbox{}}\def\@evenfoot{}
\def\sectionmark##1{}\def\subsectionmark##1{}}

\oddsidemargin=\evensidemargin
\renewcommand{\thefootnote}{\fnsymbol{footnote}}

\newcounter{sectionc}\newcounter{subsectionc}\newcounter{subsubsectionc}
\renewcommand{\section}[1] {\vspace{12pt}\addtocounter{sectionc}{1} 
\setcounter{subsectionc}{0}\setcounter{subsubsectionc}{0}\noindent 
	{\tenbf\thesectionc. #1}\par\vspace{5pt}}
\renewcommand{\subsection}[1] {\vspace{12pt}\addtocounter{subsectionc}{1} 
	\setcounter{subsubsectionc}{0}\noindent 
	{\bf\thesectionc.\thesubsectionc. {\kern1pt \bfit #1}}\par\vspace{5pt}}
\renewcommand{\subsubsection}[1] {\vspace{12pt}\addtocounter{subsubsectionc}{1}
	\noindent{\tenrm\thesectionc.\thesubsectionc.\thesubsubsectionc.
	{\kern1pt \tenit #1}}\par\vspace{5pt}}
\newcommand{\nonumsection}[1] {\vspace{12pt}\noindent{\tenbf #1}
	\par\vspace{5pt}}

\newcounter{appendixc}
\newcounter{subappendixc}[appendixc]
\newcounter{subsubappendixc}[subappendixc]
\renewcommand{\thesubappendixc}{\Alph{appendixc}.\arabic{subappendixc}}
\renewcommand{\thesubsubappendixc}
	{\Alph{appendixc}.\arabic{subappendixc}.\arabic{subsubappendixc}}

\renewcommand{\appendix}[1] {\vspace{12pt}
        \refstepcounter{appendixc}
        \setcounter{figure}{0}
        \setcounter{table}{0}
        \setcounter{lemma}{0}
        \setcounter{theorem}{0}
        \setcounter{corollary}{0}
        \setcounter{definition}{0}
        \setcounter{equation}{0}
        \renewcommand{\thefigure}{\Alph{appendixc}.\arabic{figure}}
        \renewcommand{\thetable}{\Alph{appendixc}.\arabic{table}}
        \renewcommand{\theappendixc}{\Alph{appendixc}}
        \renewcommand{\thelemma}{\Alph{appendixc}.\arabic{lemma}}
        \renewcommand{\thetheorem}{\Alph{appendixc}.\arabic{theorem}}
        \renewcommand{\thedefinition}{\Alph{appendixc}.\arabic{definition}}
        \renewcommand{\thecorollary}{\Alph{appendixc}.\arabic{corollary}}
        \renewcommand{\theequation}{\Alph{appendixc}.\arabic{equation}}
        \noindent{\tenbf Appendix \theappendixc #1}\par\vspace{5pt}}
\newcommand{\subappendix}[1] {\vspace{12pt}
        \refstepcounter{subappendixc}
        \noindent{\bf Appendix \thesubappendixc. {\kern1pt \bfit #1}}
	\par\vspace{5pt}}
\newcommand{\subsubappendix}[1] {\vspace{12pt}
        \refstepcounter{subsubappendixc}
        \noindent{\rm Appendix \thesubsubappendixc. {\kern1pt \tenit #1}}
	\par\vspace{5pt}}

\newcommand{\textlineskip}{\baselineskip=13pt}
\newcommand{\smalllineskip}{\baselineskip=10pt}

\def\eightcirc{
\begin{picture}(0,0)
\put(4.4,1.8){\circle{6.5}}
\end{picture}}
\def\eightcopyright{\eightcirc\kern2.7pt\hbox{\eightrm c}} 

\newcommand{\copyrightheading}[1]
	{\vspace*{-2.5cm}\smalllineskip{\flushleft
	 }}

\def\abstracts#1#2#3{{
	\centering{\begin{minipage}{4.5in}\baselineskip=10pt\footnotesize
	\parindent=0pt #1\par 
	\parindent=15pt #2\par
	\parindent=15pt #3
	\end{minipage}}\par}} 

\newcommand{\bibit}{\nineit}

\renewenvironment{thebibliography}[1]
	{\frenchspacing
	 \ninerm\baselineskip=11pt
	 \begin{list}{\arabic{enumi}.}
	{\usecounter{enumi}\setlength{\parsep}{0pt}
	 \setlength{\leftmargin 12.7pt}{\rightmargin 0pt} 
	 \setlength{\itemsep}{0pt} \settowidth
	{\labelwidth}{#1.}\sloppy}}{\end{list}}

\newcounter{itemlistc}
\newcounter{romanlistc}
\newcounter{alphlistc}
\newcounter{arabiclistc}

\newcommand{\fcaption}[1]{
        \refstepcounter{figure}
        \setbox\@tempboxa = \hbox{\footnotesize Fig.~\thefigure. #1}
        \ifdim \wd\@tempboxa > 5in
           {\begin{center}
        \parbox{5in}{\footnotesize\smalllineskip Fig.~\thefigure. #1}
            \end{center}}
        \else
             {\begin{center}
             {\footnotesize Fig.~\thefigure. #1}
              \end{center}}
        \fi}

\newcommand{\tcaption}[1]{
        \refstepcounter{table}
        \setbox\@tempboxa = \hbox{\footnotesize Table~\thetable. #1}
        \ifdim \wd\@tempboxa > 5in
           {\begin{center}
        \parbox{5in}{\footnotesize\smalllineskip Table~\thetable. #1}
            \end{center}}
        \else
             {\begin{center}
             {\footnotesize Table~\thetable. #1}
              \end{center}}
        \fi}

\def\@citex[#1]#2{\if@filesw\immediate\write\@auxout
	{\string\citation{#2}}\fi
\def\@citea{}\@cite{\@for\@citeb:=#2\do
	{\@citea\def\@citea{,}\@ifundefined
	{b@\@citeb}{{\bf ?}\@warning
	{Citation `\@citeb' on page \thepage \space undefined}}
	{\csname b@\@citeb\endcsname}}}{#1}}

\newif\if@cghi
\def\cite{\@cghitrue\@ifnextchar [{\@tempswatrue
	\@citex}{\@tempswafalse\@citex[]}}
\def\citelow{\@cghifalse\@ifnextchar [{\@tempswatrue
	\@citex}{\@tempswafalse\@citex[]}}
\def\@cite#1#2{{$\null^{#1}$\if@tempswa\typeout
	{IJCGA warning: optional citation argument 
	ignored: `#2'} \fi}}

\def\pmb#1{\setbox0=\hbox{#1}
	\kern-.025em\copy0\kern-\wd0
	\kern.05em\copy0\kern-\wd0
	\kern-.025em\raise.0433em\box0}

\def\fnm#1{$^{\mbox{\scriptsize #1}}$}
\def\fnt#1#2{\footnotetext{\kern-.3em
	{$^{\mbox{\scriptsize #1}}$}{#2}}}

\def\fpage#1{\begingroup
\voffset=.3in
\thispagestyle{empty}\begin{table}[b]\centerline{\footnotesize #1}
	\end{table}\endgroup}

\def\runninghead#1#2{\pagestyle{myheadings}
\markboth{{\protect\footnotesize\it{\quad #1}}\hfill}
{\hfill{\protect\footnotesize\it{#2\quad}}}}
\font\tenrm=cmr10
\font\tenit=cmti10 
\font\tenbf=cmbx10
\font\bfit=cmbxti10 at 10pt
\font\ninerm=cmr9
\font\nineit=cmti9

\font\eightrm=cmr8

\def\qed{\hbox{${\vcenter{\vbox{			
   \hrule height 0.4pt\hbox{\vrule width 0.4pt height 6pt
   \kern5pt\vrule width 0.4pt}\hrule height 0.4pt}}}$}}

\renewcommand{\thefootnote}{\fnsymbol{footnote}}	

\begin{document}

\runninghead{ H\"older Inequalities and QCD Sum-Rule Bounds on the Masses of Light Quarks} { H\"older Inequalities and QCD Sum-Rule Bounds on the Masses of Light Quarks}

\normalsize\textlineskip
\thispagestyle{empty}
\setcounter{page}{1}

\copyrightheading{}			

\vspace*{0.88truein}

\fpage{1}
\centerline{\bf H\"older Inequalities and QCD Sum-Rule Bounds on the}
\vspace*{0.035truein}
\centerline{\bf  Masses of Light Quarks}
\vspace*{0.37truein}
\centerline{\footnotesize T.G. Steele\footnote{Research funded by the 
Natural Science and Engineering Research Council of Canada (NSERC).
}}
\vspace*{0.015truein}
\centerline{\footnotesize\it Department of Physics \& Engineering Physics, University
of Saskatchewan, 116 Science Place}
\baselineskip=10pt
\centerline{\footnotesize\it Saskatoon, Saskatchewan,  S7N 5E2,
Canada}

\vspace*{0.21truein}
\abstracts{
QCD Laplace Sum-Rules must satisfy a fundamental H\"older inequality if they are to 
consistently represent an integrated hadronic spectral function. The Laplace
sum-rules of pion currents is shown to violate this inequality unless the $u$ and $d$ 
quark masses are sufficiently large, placing a lower bound on $m_u+m_d$, the $SU(2)$-invariant combination of 
the light-quark masses. 
}{}{}

\textlineskip			
\vspace*{12pt}			

\noindent
In this paper we briefly review the development 
of H\"older inequalities for QCD sum-rules\cite{sr_holder}  
and their application to obtain light-quark ($u,d$) mass bounds.\cite{holder_bounds}

Laplace sum-rules for pseudoscalar currents with quantum numbers of the pion
relate a QCD prediction $R_5\left(M^2\right)$ to the integral of the associated  hadronic 
spectral function $\rho_5(t)$
\begin{equation}
R_5\left(M^2\right)=\frac{1}{\pi}\int\limits_{t_0}^\infty \rho(t) \exp{\left(-\frac{t}{M^2}\right)}\,dt \quad ,
\label{basic_sr}
\end{equation} 
where $t_0$ is the physical threshold for the spectral function.
Since $\rho_5(t)\ge 0$, the right-hand (phenomenological side) side of 
(\ref{basic_sr}) must satisfy integral inequalities over a measure  $d\mu=\rho_5(t)\,dt$.

H\"older's inequality over a measure $d\mu$ is 
\begin{equation}
\biggl|\int_{t_1}^{t_2} f(t)g(t) d\mu \biggr|\! \le \!
\left(\int_{t_1}^{t_2} \big|f(t)\big|^ p d\mu \right)^{\frac{1}{p}}
\!\!\!\left(\int_{t_1}^{t_2} \big|g(t)\big|^q d\mu \right)^{\frac{1}{q}}
~,~
\frac{1}{p}+\frac{1}{q} =1~;~ p,~q\ge 1 \quad ,
\label{holder_ineq}
\end{equation}
which for  $p=q=2$   reduces to the familiar Schwarz inequality, implying 
that the H\"older inequality is a more general constraint. The H\"older inequality can be applied to 
Laplace sum-rules by identifying $d\mu=\rho(t)\,dt$, $\tau=1/M^2$ and defining
\begin{equation}
S_5\left(\tau\right)=\frac{1}{\pi}\int\limits_{\mu_{th}}^\infty \!\!\rho_5(t) e^{-t\tau}\,dt
\label{s5}
\end{equation}
where $\mu_{th}$ will later be identified as lying above $m_\pi^2$.  Suitable choices of 
$f(t)$ and $g(t)$ in the H\"older inequality (\ref{holder_ineq}) yield the following 
inequality  for $S_5(t)$:\cite{sr_holder} 
\begin{equation}
S_5\left(\tau+(1-\omega)\delta\tau\right)\le S_5^\omega\left(\tau\right)
S_5^{1-\omega}\left(\tau+\delta\tau\right) 
\quad ,~ \forall~ 0\le \omega\le 1\quad .
\label{s5_ineq}
\end{equation}

\pagebreak
\textheight=7.8truein
\setcounter{footnote}{0}
\renewcommand{\thefootnote}{\alph{footnote}}

The Laplace sum-rule relating QCD and hadronic physics is obtained by applying the 
Borel transform operator\cite{SVZ} $\hat B$ 
to the dispersion relation
\begin{equation}
\Pi_{5}\left(Q^2\right)=a+b Q^2
+\frac{Q^4}{\pi}\int\limits_{t_0}^\infty
\frac{ \rho_5(t)}{t^2\left(t+Q^2\right)}\, dt\quad .
\label{dispersion}
\end{equation}
The quantity  $\Pi_5\left(Q^2\right)$ is the QCD prediction for the correlation function of 
pseudoscalar
(pion) currents
\begin{gather}
\Pi_{5}\left(Q^2\right)=i\int d^4x\, e^{iq\cdot x}\left\langle O \vert T\left[ J_{5}(x) 
J_5(0)\right] \vert O \right\rangle
\label{corr_fn}\\
J_5(x)=\frac{1}{\sqrt{2}}\left(m_u+m_d\right)\left[\bar u(x)i\gamma_5u(x)-
\bar d(x)i\gamma_5d(x)\right]\quad ,
\label{current}
\end{gather} 
and the theoretically-determined quantity   $R_5\left(M^2\right)$ is obtained from  the Borel 
transform of the QCD 
correlation function.
\begin{equation}
R_{5}\left(M^2\right)=M^2\hat B\left[\Pi_{5}\left(Q^2\right)\right]
\end{equation}

  Perturbative contributions to  $R_5\left(M^2\right)$ are known up to four-loop order.\cite{chetyrkin,BNRY}  
Infinite correlation-length vacuum effects in  $R_5\left(M^2\right)$
are represented by  the (non-perturbative) QCD condensate contributions.\cite{SVZ,BNRY,Bagan}  
In addition to the QCD condensate contributions the pseudoscalar (and scalar) correlation functions
are sensitive to finite correlation-length vacuum effects described by direct instantons in the instanton 
liquid model.\cite{EVS}   The total result for $R_5\left(M^2\right)$ to leading order 
in the light-quark masses is\cite{holder_bounds}
\begin{equation}
\begin{split}
{ R}_5\left(M^2\right)=&\frac{3m^2M^4}{8\pi^2}\left(   
1+4.821098 \frac{\alpha}{\pi}+21.97646\left(\frac{\alpha}{\pi}\right)^2+53.14179\left(\frac{\alpha}{\pi}\right)^3
\right)
\\
& +m^2\left(
-\langle m\bar q q\rangle 
+\frac{1}{8\pi}\langle \alpha G^2\rangle
+\frac{\pi\langle{\cal O}_6\rangle}{4M^2}
\right)
\\
& +m^2
{3\rho_c^2 M^6\over{8 \pi^2}} e^{-\rho_c^2M^2/2 }
\left[   
  K_0\left( {\rho_c^2M^2/2} \right) +
       K_1\left( {\rho_c^2M^2/2} \right)
\right]\quad ,
\end{split}
\label{R_5}
\end{equation} 
where $\alpha$ and  $m=\left(m_u+m_d\right)/2$ are the $\overline{MS}$ 
running coupling and quark masses at the scale $M$, and $\rho_c$ represents the instanton size in the 
instanton liquid model.
Note that all the theoretical contributions are proportional to  $m^2$, demonstrating that the quark mass sets
the scale of the pseudoscalar channel.   Higher-loop perturbative contributions in (\ref{R_5}) are significant, 
and effectively enhance the quark mass with increasing loop order.  

To employ the H\"older inequality (\ref{s5_ineq}) we separate out the pion pole by setting $\mu_{th}=9m_\pi^2$ 
in (\ref{s5}).
\begin{equation}
S_5\left(M^2\right)=R_5\left(M^2\right)-2f_\pi^2m_\pi^4=\int\limits_{9m_\pi^2}^\infty \rho_5(t)
 e^{-t\tau}\,dt
\label{s5_fin}
\end{equation}
Lower bounds on the quark mass $m$ can then be obtained by finding the minimum value of $m$ for which the
H\"older inequality (\ref{s5_ineq}) is satisfied. 
Introducing further phenomenological 
contributions ({\it e.g.} three-pion continuum) would tend to give a larger mass bound.  However, 
 if only the  pion pole is separated out, then the analysis is not subject to uncertainties introduced by the
 phenomenological model. 

Standard values of the QCD parameters are employed in the inequality analysis of (\ref{s5_ineq}), 
and we  use
$\delta\tau\lesssim 0.1\,{\rm GeV^{-2}}$   for which this analysis becomes local (depending only on the 
Borel scale $M$).\cite{sr_holder,holder_bounds}  
Validity of QCD predictions at the $\tau$ mass is evidenced by the 
analysis of the $\tau$ hadronic width, hadronic contributions to $\alpha_{EM}\left(M_Z\right)$ and the 
muon anomalous magnetic moment,\cite{braaten_davier} so 
we impose the inequality (\ref{s5_ineq}) at the $\tau$ mass  $M=M_\tau$. 
The resulting H\"older inequality bound on the $\overline{MS}$ quark masses scaled to $1.0\,{\rm GeV}$ 
is\cite{holder_bounds}
\begin{equation}
m(1\,{\rm GeV})=\frac{1}{2}\left[ m_u(1\,{\rm GeV})+m_d(1\,{\rm GeV})\right]\ge 3\,{\rm MeV}
\label{final_bound}
\end{equation}
For comparison with other determinations of the light quark masses,  this result has been converted to 
\begin{equation}
m(2\,{\rm GeV})=\frac{1}{2}\left[ m_u(2\,{\rm GeV})+m_d(2\,{\rm GeV})\right]\ge 2.1\,{\rm MeV}
\label{final_pdg_bound}
\end{equation}
by the Particle Data Group.\cite{PDG}

The theoretical uncertainties in the quark mass bound (\ref{final_bound}) from the QCD parameters and 
(estimated) higher-order perturbative effects are less than 5\%, and the result (\ref{final_bound}) 
is the absolute lowest bound resulting from  the  uncertainty analysis.\cite{holder_bounds}  
 for $M\gtrsim M_\tau$ the theoretical uncertainties in the mass bound are  
$\lesssim 0.1 \, {\rm MeV}$.  Thus we have not extracted   mass bounds below  
the energy scale $M\approx M_\tau$  at which theoretical 
uncertainties first reach a non-negligible level.
Finally, compared with the
positivity inequality $S_5\left(M^2\right)\ge 0$ (as first used to obtain quark mass bounds from QCD 
sum-rules\cite{BNRY})
the H\"older inequality leads to  quark mass bounds larger by a factor of 2 for
identical theoretical and phenomenological inputs at $M=M_\tau$, demonstrating that the H\"older inequality 
provides  stringent constraints on the quark mass.

\nonumsection{References}


\begin{thebibliography}{99}


\bibitem{sr_holder}
  M.\ Benmerrouche, G.\ Orlandini, T.G.\ Steele,
{\bibit Phys.\ Lett.\ } {B356} (1995) 573.

\bibitem{holder_bounds}
T.G.\ Steele, K.\ Kostuik, J.\ Kwan, 
{\bibit Phys.\ Lett.\ }  B451 (1999) 201.

\bibitem{SVZ} M.A.\ Shifman, A.I.\ Vainshtein, V.I.\ Zakharov,  {\bibit Nucl.\ Phys.\ } {B147} (1979) 385.


\bibitem{chetyrkin} K.G.\ Chetyrkin, {\bibit Phys.\ Lett.\ } { B390} (1997) 309.

\bibitem{BNRY}
C.\ Becchi, S.\ Narison, E.\ de Rafael, F.J.\ Yndurain,
{\bibit Z. Phys. }{ C8} (1981) {335}.


\bibitem{Bagan}E.\ Bagan, J.I.\ Latorre, P.\ Pascual,
{\bibit Z.\ Phys.\ } {C32} (1986) 43.

\bibitem{EVS}
E.V.\ Shuryak, {\bibit Nucl.\ Phys.\ } {B214} (1983) {237}.

\bibitem{braaten_davier} E.\ Braaten, S.\ Narison, A.\ Pich, {\bibit Nucl.\ Phys.\ } { B373} (1992) 581; 
M.\ Davier, A.\ H\"ocker, {\bibit Phys.\ Lett.\ } {B435} (1998) 427.

\bibitem{PDG} D.E.\ Groom {\it et al.} (Particle Data Group), {\bibit Eur.\ Phys.\ Jour.\ } C15 (2000) 1.

\end{thebibliography}
\end{document}